\documentclass[doublespacing]{elsart}

\usepackage{latexsym}
\usepackage{amstext,amssymb,amsfonts}
\usepackage{epsfig}

\begin{document}

\begin{frontmatter}



\title{Transport of multiple users in complex networks}

\author[1]{Shai Carmi}
\author[2]{Zhenhua Wu}
\author[3]{Eduardo L\'{o}pez}
\author[1,2]{Shlomo Havlin}
\author[2]{H. Eugene Stanley}
\address[1]{Minerva Center \& Department of Physics, Bar-Ilan University, Ramat Gan 52900, Israel}
\address[2]{Center for Polymer Studies, Boston University, Boston, MA 02215 USA}
\address[3]{Theoretical Division, Los Alamos National Laboratory, Mail Stop B258, Los Alamos, NM 87545 USA}

\begin{abstract}
We study the transport properties of model networks such as scale-free 
and Erd\H{o}s-R\'{e}nyi networks as well as a real network. 
We consider the conductance $G$ between two arbitrarily chosen nodes
where each link has the same unit resistance. Our theoretical
analysis for scale-free networks predicts a broad range
of values of $G$, with a power-law tail distribution $\Phi_{\rm SF}(G)\sim G^{-g_G}$, 
where $g_G=2\lambda -1$, and $\lambda$ is the decay exponent for the scale-free 
network degree distribution. We confirm our predictions by large scale simulations. 
The power-law tail in $\Phi_{\rm SF}(G)$ leads to large values of $G$, thereby
significantly improving the transport in scale-free networks, compared
to Erd\H{o}s-R\'{e}nyi networks where the tail of the conductivity
distribution decays exponentially. We develop a simple physical picture
of the transport to account for the results.
We study another model for transport, the \emph{max-flow} model,
where conductance is defined as the number of link-independent paths
between the two nodes, and find that a similar picture holds.
The effects of distance on the value of conductance are considered for
both models, and some differences emerge. 
We then extend our study to the case of multiple sources, where the transport
is define between two \emph{groups} of nodes. We find a fundamental difference
between the two forms of flow when considering the quality of the transport with respect 
to the number of sources, and find an optimal number of sources, or users,
for the max-flow case. A qualitative (and partially quantitative) explanation
is also given.

\end{abstract}

\end{frontmatter}

\section{Introduction}

Transport in many random structures is ``anomalous,'' i.e.,
fundamentally different from that in regular space
\cite{havlin-ben-avraham,ben-avraham-havlin,bunde-havlin}.
The anomaly is due to the random substrate on which transport is
constrained to take place.
Random structures are found in many places in the real world, from oil
reservoirs to the Internet, making the study of anomalous transport
properties a far-reaching field.  In this problem, it is paramount to
relate the structural properties of the medium with the transport
properties.

An important and recent example of random substrates is that of complex
networks.  Research on this topic has uncovered their importance for
real-world problems as diverse as the World Wide Web and the Internet to
cellular networks and sexual-partner networks \cite{rev-Albert}.
Networks describe also economic systems such as financial markets and banks systems 
\cite{Bonnano,Kertesz,Inaoka}. Transport of goods and information in such networks is of much interest.

\begin{figure}[t]
\begin{center}
\epsfig{file=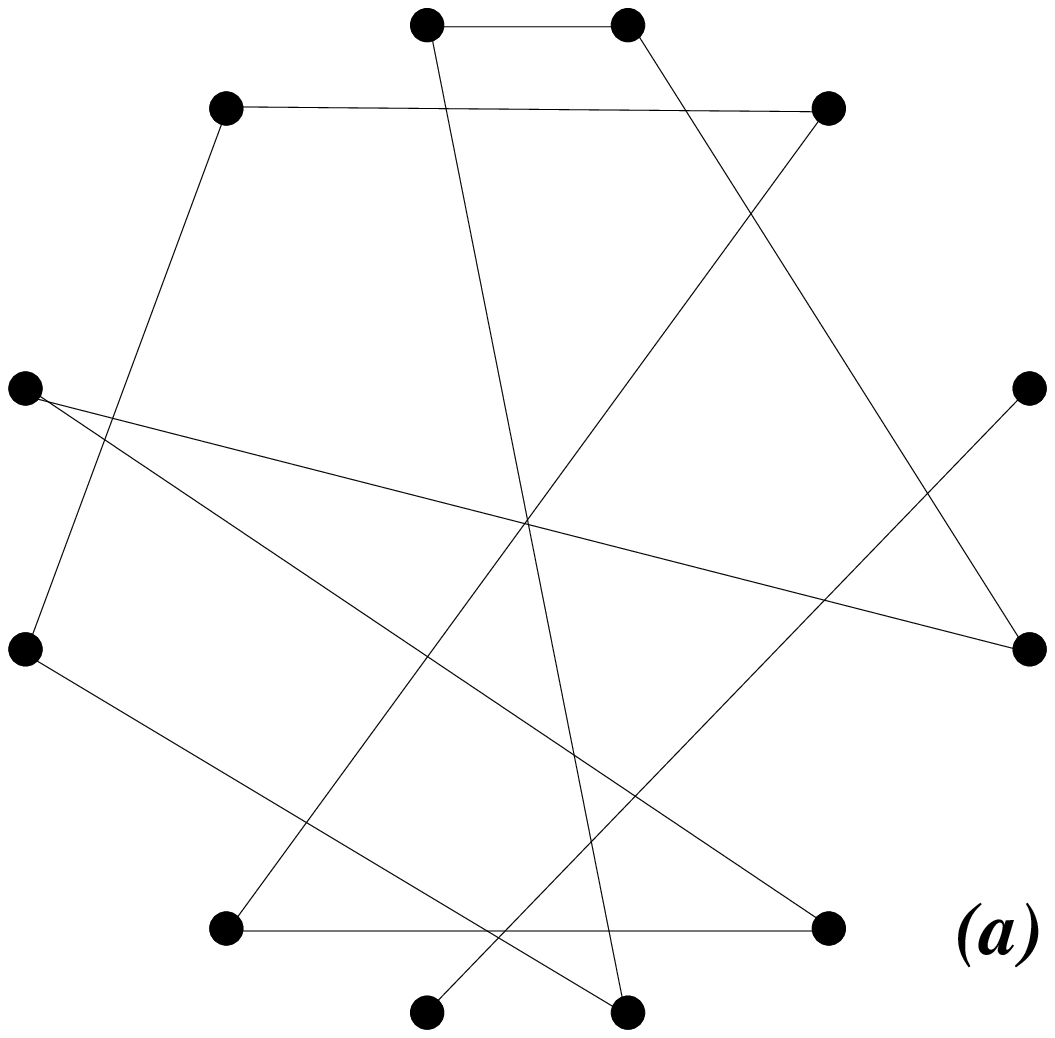,height=3.6cm,width=3.8cm}
\hspace{0.5cm}
\epsfig{file=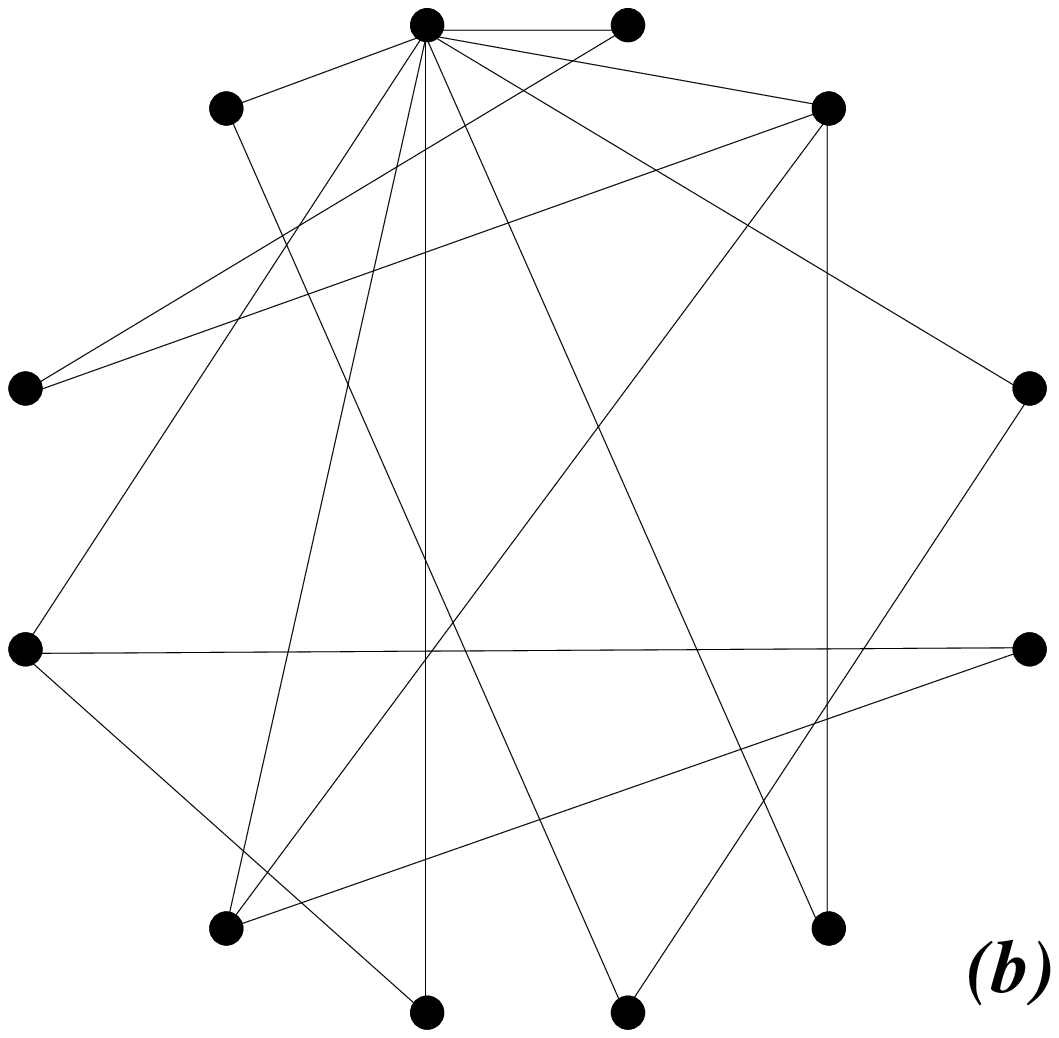,height=3.6cm,width=3.8cm}
\end{center}
\caption{(a) Schematic of an Erd\H{o}s-R\'{e}nyi network of $N=12$ and
$p=1/6$. Note that in this example ten nodes
have $k=2$ connections, and two nodes have $k=1$ connections.
This illustrates the fact that for Erd\H{o}s-R\'{e}nyi networks,
the range of values of degree is very narrow, typically close to
$\overline{k}$.
(b) Schematic of a scale-free network of $N=12$, $k_{\rm min}=2$ and
$\lambda\approx 2$. We note the presence of a hub with $k_{\rm max}=8$
which is connected to many of the other nodes of the network.}
\label{ER_diag}
\end{figure}

Two distinct models describe the two limiting cases for the structure
of the complex networks. The first of these is the classic Erd\H{o}s-R\'{e}nyi
model of random networks \cite{ER}, for which sites are connected with a
link with probability $p$ and are disconnected (no link) with probability
$1-p$ (see Fig. \ref{ER_diag}(a)).
In this case the degree distribution $P(k)$,
the probability of a node to have $k$ connections, is a Poisson
\begin{equation}
P(k) = \frac{\left(\overline{k}\right)^k e^{-\overline{k}}}{k!},
\label{Poisson}
\end{equation}
where $\overline{k}\equiv\sum_{k=1}^{\infty} kP(k)$ is the average
degree of the network.  Mathematicians discovered critical
phenomena through this model. For instance, just as in percolation
on lattices, there is a critical value $p=p_c$ above which the
largest connected component of the network has a mass that scales
with the system size $N$, but below $p_c$, there are only small
clusters of the order of $\log N$. At $p=p_c$, the size of the largest cluster
is of order of $N^{2/3}$.Another characteristic of an
Erd\H{o}s-R\'{e}nyi network is its ``small-world'' property which
means that the average distance $d$ (or diameter) between all
pairs of nodes of the network scales as $\log N$ \cite{Bollobas}.
The other model, recently identified as the characterizing
topological structure of many real world systems, is the
Barab\'{a}si-Albert scale-free network and its extensions
\cite{scale-Barabasi,dyn-network,Simon}, characterized by a
scale-free degree distribution:
\begin{equation}
P(k)\sim k^{-\lambda}\qquad [k_{\rm min}\le k\le k_{\rm max}],
\label{degree}
\end{equation}
The cutoff value $k_{\rm min}$ represents the minimum allowed value of
$k$ on the network ($k_{\rm min}=2$ here), and $k_{\rm max}\equiv k_{\rm
min}N^{1/(\lambda-1)}$, the typical maximum degree of a network with $N$
nodes \cite{Cohen,netwcomment}.  The scale-free feature allows a network
to have some nodes with a large number of links (``hubs''), unlike the
case for the Erd\H{o}s-R\'{e}nyi model of random networks \cite{ER,Bollobas} 
(See Fig.~\ref{ER_diag}(b)).
Scale-free networks with $\lambda >3$ have $d\sim \log N$, while for
$2<\lambda <3$ they are ``ultra-small-world'' since the diameter
scales as $d\sim \log\log N$ \cite{Cohen}.

Here we extend our recent study of transport in complex networks
\cite{Lopez-PRL,book-chap}.  We find that for scale-free networks with
$\lambda\ge 2$, transport properties characterized by conductance display a
power-law tail distribution that is related to the degree distribution
$P(k)$.  The origin of this power-law tail is due to pairs of nodes of high
degree which have high conductance.  Thus, transport in scale-free networks
is better because of the presence of large degree nodes (hubs) that carry
much of the traffic, whereas Erd\H{o}s-R\'{e}nyi networks lack hubs and the
transport properties are controlled mainly by the average degree
$\overline{k}$ \cite{Bollobas,Grimmett-Kesten}.  We present a simple physical
picture of the transport properties and test it through simulations.  We also
study a form of frictionless transport, in which transport is measured by the
number of independent paths between source and destination. These later
results are in part similar to those in \cite{Lee}.  We test our findings on
a real network, a recent map of the Internet.  In addition, we study the
properties of the transport where several sources and sinks are involved. We
find a principal difference between the two forms of transport mentioned
above, and find an optimal number of sources in the frictionless case.

The paper is structured as follows. Section \ref{transport}
concentrates on the numerical calculation of the electrical
conductance of networks. In Sec. \ref{theory_section} a simple physical
picture gives a theoretical explanation of the results.
Section \ref{frictionless} deals with the number of
link-independent paths as a form of frictionless transport. 
In Section \ref{multiple} we extend the study of frictionless transport
to the case of multiple sources.
Finally, in Sec. \ref{summary} we present the conclusions and summarize the
results in a coherent picture.

\section{Transport in complex networks}
\label{transport}

Most of the work done so far regarding complex networks has concentrated
on static topological properties or on models for their growth
\cite{rev-Albert,Cohen,dyn-network,Toroczkai}.  Transport features have
not been extensively studied with the exception of random walks on
specific complex networks \cite{Noh,Sood,Gallos}. Transport
properties are important because they contain information about network function
\cite{dyn-topology}.  Here we study the electrical conductance $G$
between two nodes $A$ and $B$ of Erd\H{o}s-R\'{e}nyi and scale-free
networks when a potential difference is imposed between them.  We assume
that all the links have equal resistances of unit value
\cite{comm-struc}.

We construct the Erd\H{o}s-R\'{e}nyi and scale-free model networks in a standard
manner \cite{Lopez-PRL,Molloy-Reed}.
We calculate the conductance $G$ of the network between two nodes $A$
and $B$ using the Kirchhoff method \cite{Kirchhoff}, where entering and
exiting potentials are fixed to $V_A=1$ and $V_B=0$.

\begin{figure}[t]
\begin{center}
\epsfig{file=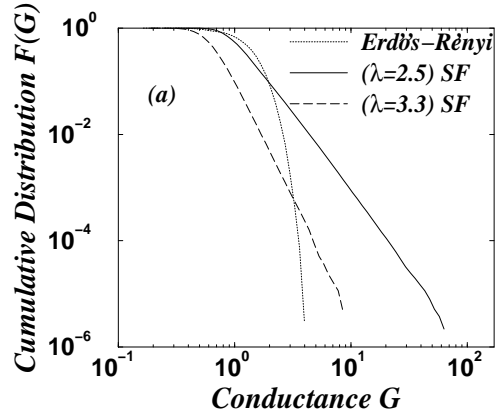,height=6cm,width=6cm}
\end{center}
\caption{Comparison for networks with $N=8000$ nodes between the
  cumulative distribution functions of conductance for the Erd\H{o}s-R\'{e}nyi and the
  scale-free cases (with $\lambda=2.5$ and 3.3). Each curve represents
  the cumulative distribution $F(G)$ vs. $G$. The simulations have at
  least $10^6$ realizations.}
\label{FG_lamb2.5-3.3-ER_N8000}
\end{figure}

First, we analyze the probability density function (pdf) $\Phi(G)$ which
comes from $\Phi(G)dG$, the probability that two nodes on the network
have conductance between $G$ and $G+dG$.  To this end, we introduce the
cumulative distribution $F(G)\equiv\int_{G}^\infty \Phi(G')dG'$, shown
in Fig.~\ref{FG_lamb2.5-3.3-ER_N8000} for the Erd\H{o}s-R\'{e}nyi and
scale-free ($\lambda=2.5$ and $\lambda=3.3$, with $k_{\rm min}=2$)
cases.  We use the notation $\Phi_{\rm SF}(G)$ and $F_{\rm SF}(G)$ for
scale-free, and $\Phi_{\rm ER}(G)$ and $F_{\rm ER}(G)$ for
Erd\H{o}s-R\'{e}nyi.  The function $F_{\rm SF}(G)$ for both
$\lambda=2.5$ and 3.3 exhibits a tail region well fit by the power law
\begin{equation}
F_{\rm SF}(G)\sim G^{-(g_G-1)},
\end{equation}
and the exponent $(g_G-1)$ increases with $\lambda$.  In contrast,
$F_{\rm ER}(G)$ decreases exponentially with $G$.

We next study the origin of the large values of $G$ in scale-free
networks and obtain an analytical relation between $\lambda$ and $g_G$.
Larger values of $G$ require the presence of many parallel paths, which
we hypothesize arise from the high degree nodes.  Thus, we expect that
if either of the degrees $k_A$ or $k_B$ of the entering and exiting
nodes is small (e.g. $k_A>k_B$), the conductance $G$ between $A$ and $B$
is small since there are at most $k$ different parallel branches coming
out of a node with degree $k$.  Thus, a small value of $k$ implies a
small number of possible parallel branches, and therefore a small value
of $G$.  To observe large $G$ values, it is therefore necessary that
both $k_A$ and $k_B$ be large.

\begin{figure}[t]
\begin{center}
\epsfig{file=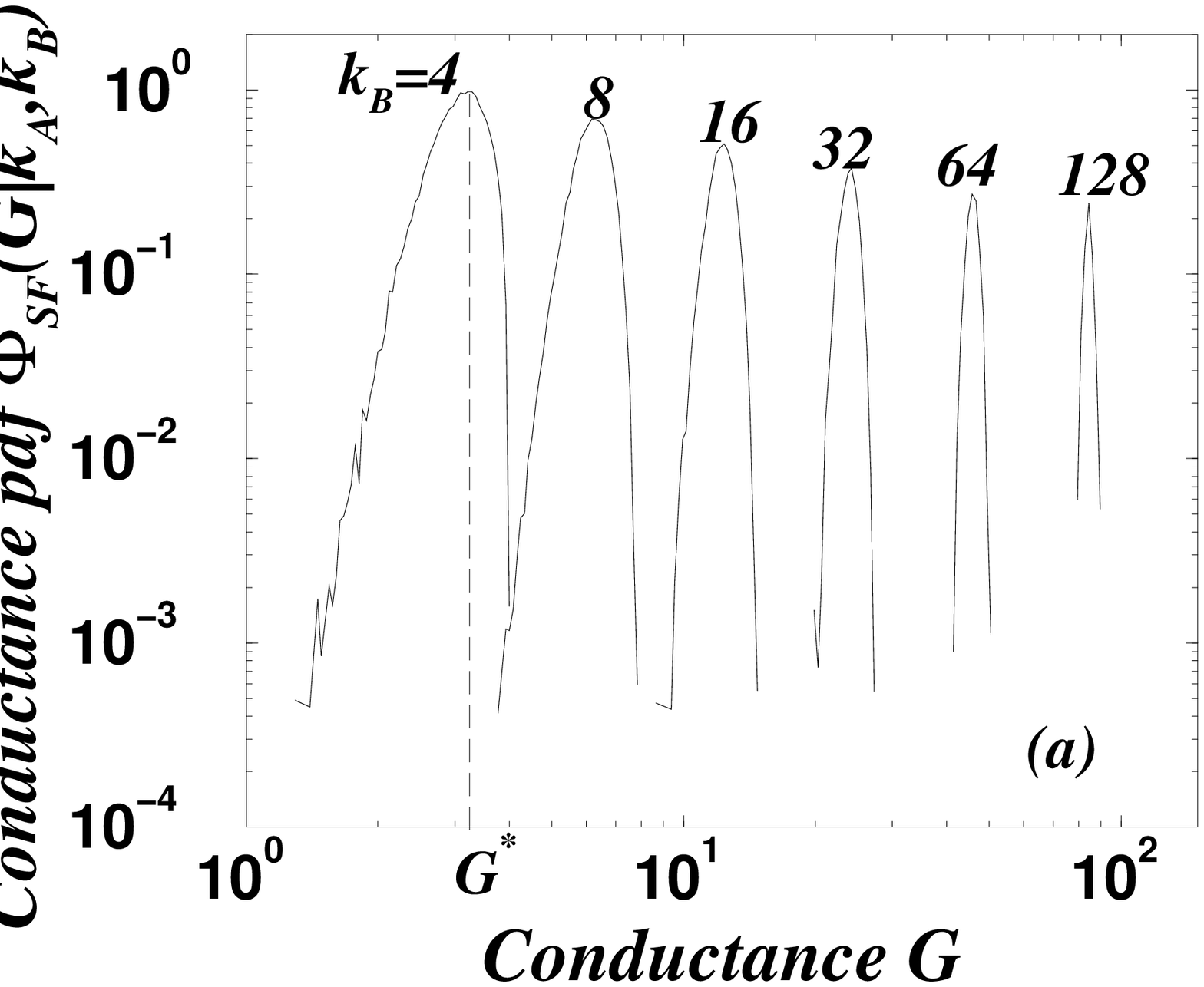,height=6.2cm,width=6cm}
\epsfig{file=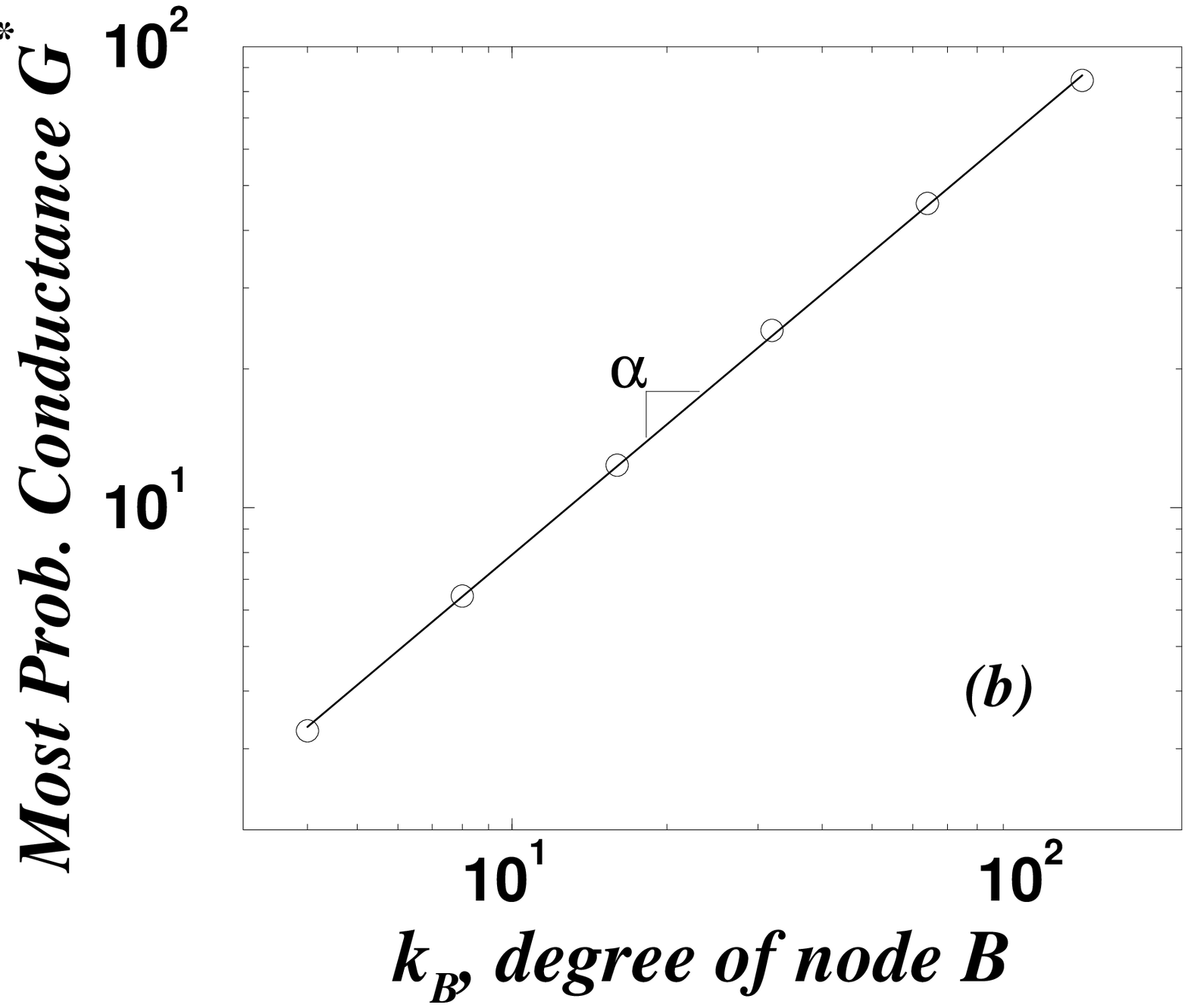,height=6cm,width=6cm}
\end{center}
\caption{(a) The pdf $\Phi_{\rm SF}(G|k_A,k_B)$ vs. $G$ for $N=8000$,
     $\lambda=2.5$ and $k_A=750$ ($k_A$ is close to the typical maximum
     degree $k_{\rm max}=800$ for $N=8000$). (b) Most probable values $G^*$,
     estimated from the maximum of the distributions in (a), as a function of
     the degree $k_B$. The data support a power law behavior $G^*\sim
     k_B^{\alpha}$ with $\alpha=0.96\pm 0.05$.}
\label{PGka750_kb4-128_N8000_ab7_lamb2.5}
\end{figure}

We test this hypothesis by large scale computer simulations of the
conditional pdf $\Phi_{\rm SF}(G|k_A,k_B)$ for specific values of the
entering and exiting node degrees $k_A$ and $k_B$.  Consider first
$k_B\ll k_A$, and the effect of increasing $k_B$, with $k_A$ fixed.  We
find that $\Phi_{\rm SF}(G|k_A,k_B)$ is narrowly peaked
(Fig.~\ref{PGka750_kb4-128_N8000_ab7_lamb2.5}(a)) so that it is well
characterized by $G^*$, the value of $G$ when $\Phi_{\rm SF}$ is a
maximum.  We find similar results for Erd\H{o}s-R\'{e}nyi networks.
Further, for increasing $k_B$, we find
[Fig.~\ref{PGka750_kb4-128_N8000_ab7_lamb2.5}(b)] $G^*$ increases as
$G^*\sim k_B^{\alpha}$, with $\alpha=0.96\pm 0.05$ consistent with the
possibility that as $N\rightarrow\infty$, $\alpha=1$ which we assume
henceforth.

For the case of $k_B\gtrsim k_A$, $G^*$ increases less fast than
$k_B$, as can be seen in Fig.~\ref{G_over_kb_vs_ka_over_kb_combi_N8000_lamb2.5_ab7_m2}
where we plot $G^*/k_B$ against the scaled degree $x\equiv k_A/k_B$. The
collapse of $G^*/k_B$ for different values of
$k_A$ and $k_B$ indicates that $G^*$ scales as
\begin{equation}
G^*\sim k_Bf\left(\frac{k_A}{k_B}\right).
\label{G_max_scaled}
\end{equation}
Below we study the possible origin of this function.

\section{Transport backbone picture}
\label{theory_section}

\begin{figure}[t]
\begin{center}
\epsfig{file=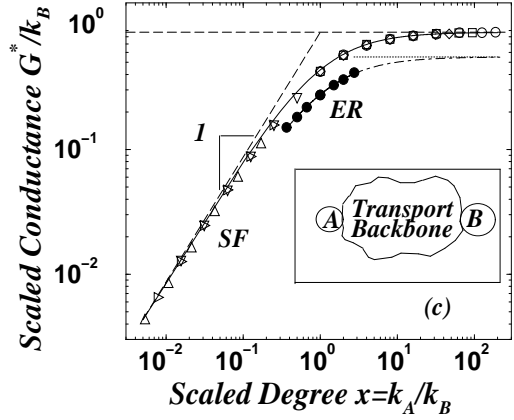,height=6cm,width=6cm}
\end{center}
\caption{ Scaled most probable
     conductance $G^*/k_B$ vs. scaled degree $x\equiv k_A/k_B$ for
     system size $N=8000$ and $\lambda=2.5$, for several values of $k_A$
     and $k_B$: $\Box$ ($k_A=8$, $8\le k_B\le 750$), $\diamondsuit$
     ($k_A=16$, $16\le k_B\le 750$), $\bigtriangleup$ ($k_A=750$, $4\le
     k_B\le 128$), $\bigcirc$ ($k_B=4$, $4\le k_A\le 750$),
     $\bigtriangledown$ ($k_B=256$, $256\le k_A\le 750$), and
     $\triangleright$ ($k_B=500$, $4\le k_A\le 128$).  The curve
     crossing the symbols is the predicted function
     $G^*/k_B=f(x)=cx/(1+x)$ obtained from Eq.~(\ref{e4b}).  We also
     show $G^*/k_B$ vs. scaled degree $x\equiv k_A/k_B$ for
     Erd\H{o}s-R\'{e}nyi networks with $\overline{k}=2.92$, $4\le k_A\le
     11$ and $k_B=4$ (symbol $\bullet$). The curve crossing the symbols
     represents the theoretical result according to Eq.~(\ref{e4b}), and
     an extension of this line to represent the limiting value of
     $G^*/k_B$ (dotted-dashed line).  The probability of observing $k_A>11$
     is extremely small in Erd\H{o}s-R\'{e}nyi networks, and thus we are
     unable to obtain significant statistics. The scaling function $f(x)$,
     as seen here, exhibits a crossover from a linear behavior to the
     constant $c$ ($c=0.87\pm0.02$ for scale-free networks, horizontal
     dashed line, and $c=0.55\pm0.01$ for Erd\H{o}s-R\'{e}nyi, dotted
     line).  The inset shows a schematic of the ``transport backbone''
     picture, where the circles labeled $A$ and $B$ denote nodes $A$ and
     $B$ and their associated links which do not belong to the
     ``transport backbone''.}
\label{G_over_kb_vs_ka_over_kb_combi_N8000_lamb2.5_ab7_m2}
\end{figure}

The behavior of the scaling function $f(x)$ can be interpreted using the
following simplified ``transport backbone'' picture
[Fig.~\ref{G_over_kb_vs_ka_over_kb_combi_N8000_lamb2.5_ab7_m2} inset], for which the
effective conductance $G$ between nodes $A$ and $B$ satisfies
\begin{equation}
{1\over G}={1\over G_A}+{1\over G_{tb}}+{1\over G_B},
\label{e4a}
\end{equation}
where $1/G_{tb}$ is the resistance of the ``transport backbone'' while
$1/G_A$ (and $1/G_B$) are the resistances of the set of links near node
$A$ (and node $B$) not belonging to the ``transport backbone''. It is
plausible that $G_A$ is linear in $k_A$, so we can write
$G_A=ck_A$. Since node $B$ is equivalent to node $A$, we expect
$G_B=ck_B$. Hence
\begin{equation}
G= \frac{1}{1/ck_A +1/ck_B+1/G_{tb}}
=k_B\frac{ck_A/k_B}{1+k_A/k_B+ck_A/G_{tb}},
\label{Gh}
\end{equation}
so the scaling function defined in Eq.~(\ref{G_max_scaled}) is
\begin{equation}
f(x)={cx\over 1+x+ck_A/G_{tb}}\approx{cx\over1+x}.
\label{e4b}
\end{equation}
The second equality follows if there are many parallel paths on the
``transport backbone'' so that $1/G_{tb}\ll 1/ck_A$.  The
prediction (\ref{e4b}) is plotted in
Fig.~\ref{G_over_kb_vs_ka_over_kb_combi_N8000_lamb2.5_ab7_m2}
for both scale-free and Erd\H{o}s-R\'{e}nyi networks
and the agreement with
the simulations supports the approximate validity of the transport
backbone picture of conductance in scale-free and Erd\H{o}s-R\'{e}nyi
networks.

\begin{figure}[t]
\begin{center}
\epsfig{file=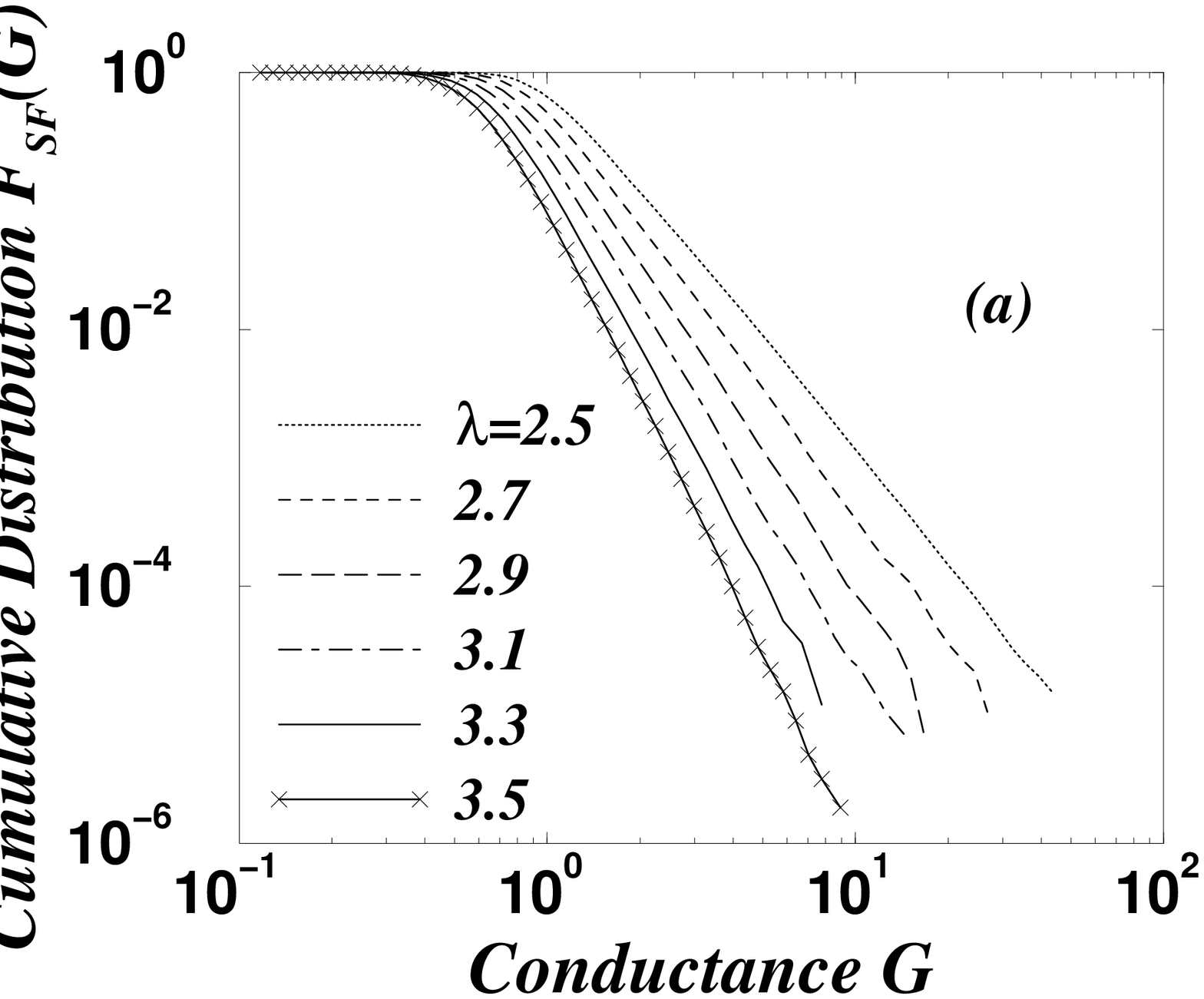,height=6.6cm,width=6cm}
\epsfig{file=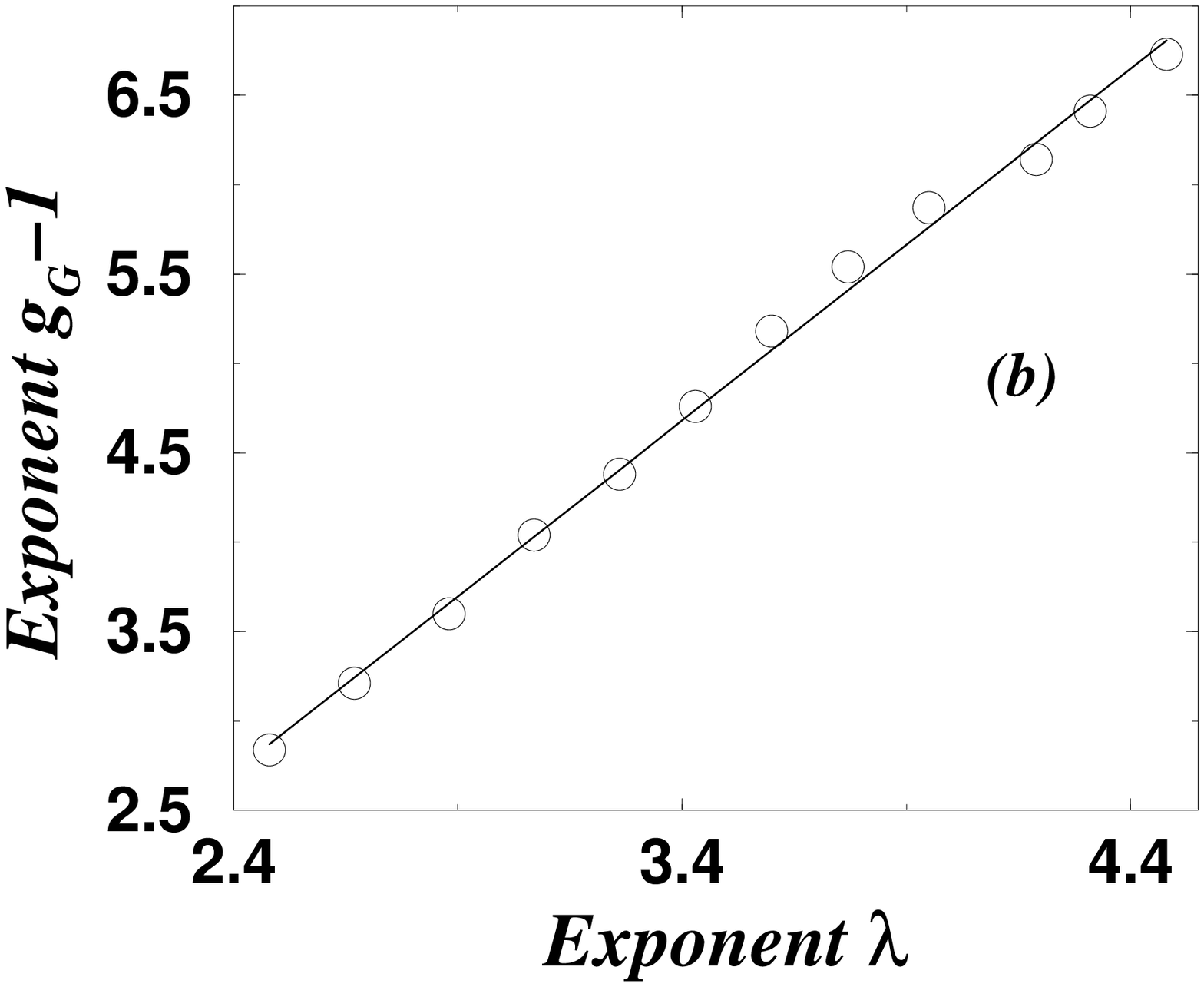,height=5.8cm,width=6cm}
\end{center}
\caption{(a) Simulation results for the cumulative distribution $F_{\rm
     SF}(G)$ for $\lambda$ between 2.5 and 3.5, consistent with the
     power law $F_{\rm SF}\sim G^{-(g_G-1)}$ (cf. Eq.~(\ref{PhiG})),
     showing the progressive change of the slope $g_G-1$. (b) The
     exponent $g_G-1$ from simulations (circles) with $2.5<\lambda<4.5$;
     shown also is a least
     square fit $g_G -1=(1.97\pm 0.04)\lambda -(2.01\pm 0.13)$,
     consistent with the predicted expression $g_G-1=2\lambda -2$
     [cf. Eq.~(\ref{PhiG})].}
\label{FG_G_lamb2.5-3.5_8000}
\end{figure}

Within this ``transport backbone'' picture, we can analytically
calculate $F_{\rm SF}(G)$.
The key insight necessary for this calculation is that $G^*\sim k_B$, when
$k_B\leq k_A$, and we assume that $G\sim k_B$ is also
valid given the narrow shape of $\Phi_{\rm SF}(G|k_A,k_B)$.
This implies that the probability of observing
conductance $G$ is related to $k_B$ through
$\Phi_{\rm SF}(G)dG\sim M(k_B)dk_B$, where $M(k_B)$ is the probability that,
when nodes $A$ and $B$ are chosen at random, $k_B$ is the minimum
degree. This can be calculated analytically through
\begin{equation}
\label{k_min_eq}
M(k_B)\sim P(k_B)\int^{k_{\rm max}}_{k_B} P(k_A) dk_A
\end{equation}
Performing the integration 
we obtain for $G<G_{\rm max}$
\begin{equation}
\Phi_{\rm SF}(G)\sim G^{-g_G} \qquad [g_G=2\lambda -1].
\label{PhiG}
\end{equation}
Hence, for $F_{\rm SF}(G)$, we have
$F_{\rm SF}(G)\sim G^{-(2\lambda -2)}$.
To test this prediction, we perform simulations for scale-free networks
and calculate the values of $g_G-1$ from the slope of a log-log plot of
the cumulative distribution $F_{\rm SF}(G)$.  From
Fig.~\ref{FG_G_lamb2.5-3.5_8000}(b) we find that
\begin{equation}
g_G -1=(1.97\pm 0.04)\lambda -(2.01\pm 0.13).
\label{g_G_measured}
\end{equation}
Thus, the measured slopes are consistent with the theoretical values
predicted by Eq.~(\ref{PhiG}).

\section{Number of link-independent paths: transport without friction}
\label{frictionless}

In many systems, it is the nature of the transport process that
the particles flowing through the network links experience no
friction. For example, this is the case in an electrical system
made of super-conductors \cite{kirk}, or in the case of water
flow along pipes, if frictional effects are minor. Other examples
are flow of cars along traffic routes, and perhaps most important,
the transport of information in communication networks. Common to
all these processes is that, the quality of the transport is
determined by the number of link-independent paths leading from the source
to the destination (and the capacity of each path), and not by the
length of each path (as is the case for simple electrical
conductance). In this section, we focus on non-weighted networks, and
define the conductance, as the number of link-independent paths 
between a given source and destination (sink) $A$ and $B$. 
We name this transport process as the \emph{max-flow model}, 
and denote the conductance as $G_{\rm MF}$.
Fast algorithms for solving the max-flow problem, given a network
and a pair $(A,B)$ are well known within the computer science
community \cite{maxflow}. We apply those methods to random
scale-free and Erd\H{o}s-R\'{e}nyi networks, and observe
similarities and differences from the electrical conductance
transport model. Max-flow analysis has been applied recently for
complex networks in general \cite{Lee,Wu}, and for the Internet in
particular \cite{internet}, where it was used as a significant
tool in the structural analysis of the underlying network.

\begin{figure}[t]
\begin{center}
\epsfig{file=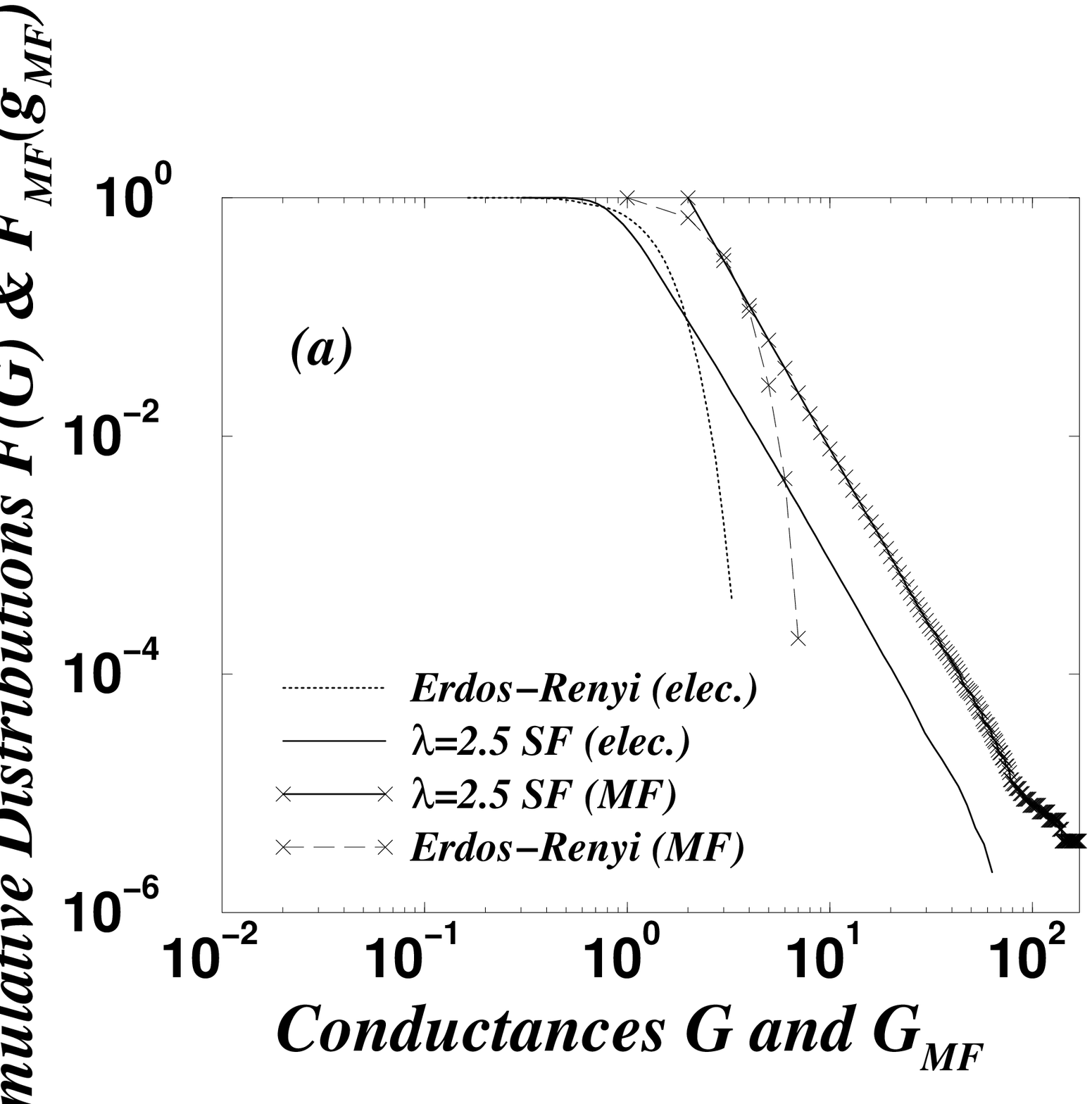,height=6cm,width=6cm}
\epsfig{file=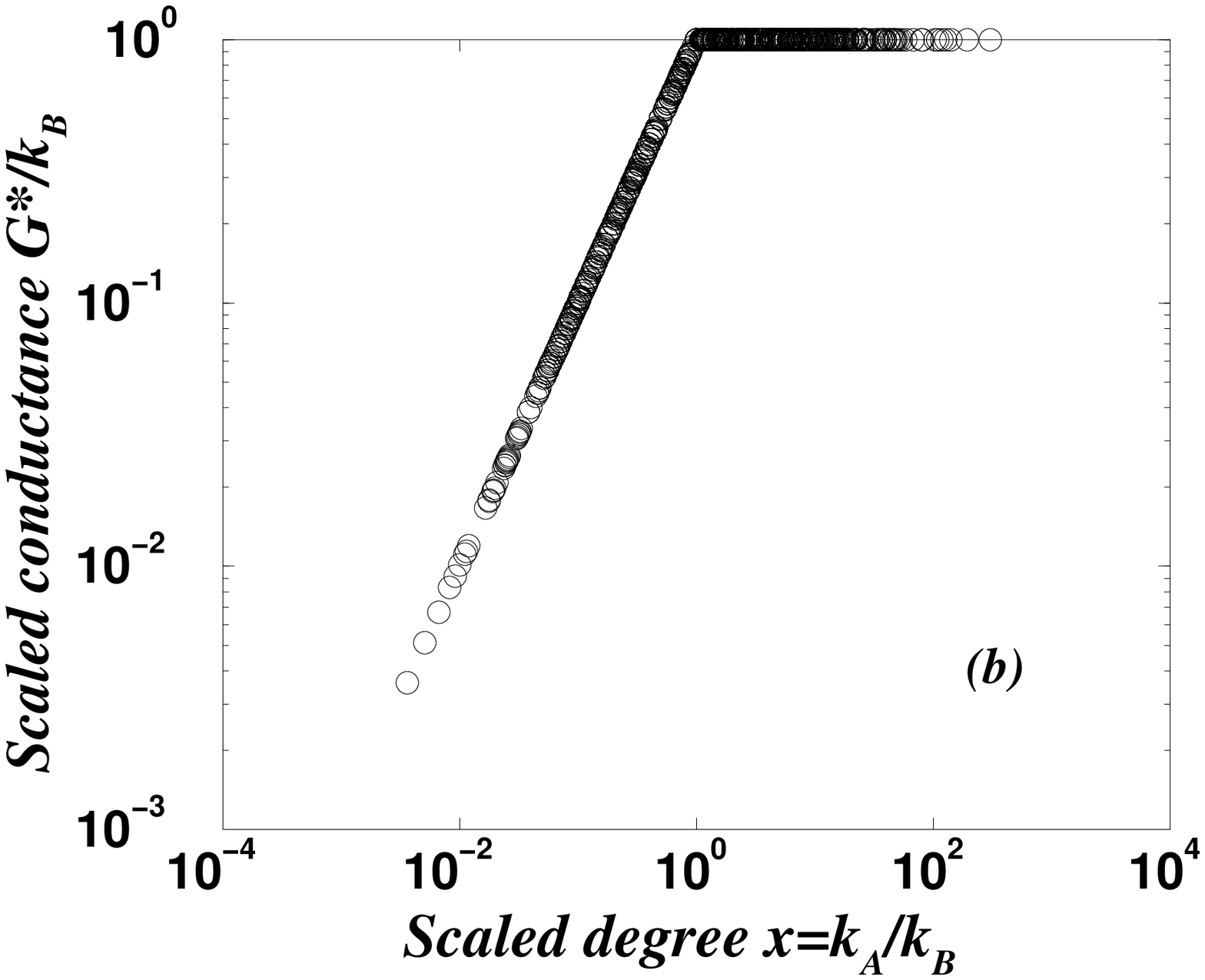,height=6cm,width=6.8cm}
\end{center}
\caption{(a) Cumulative distribution of link-independent paths
(conductance) $F_{\rm MF}(G_{\rm MF})$ vs. $G_{\rm MF}$ compared
with the electrical conductance distributions taken from
Fig.~\ref{FG_lamb2.5-3.3-ER_N8000}. We see that the scaling is
indeed the same for both models, but the proportionality constant
of $F_{\rm MF}(G_{\rm MF})$ vs. $G_{\rm MF}$ is larger for the
frictionless problem. (b) Scaled number of
independent paths $G_{\rm MF}/k_B$ as a function of the scaled
degree $k_A/k_B$ for scale-free networks of $N=8000$,
$\lambda=2.5$ and $k_{\rm min}=2$. The behavior is sharp, and shows
how $G_{\rm MF}$ is a function of only the minimum $k$.}
\label{flow_cumulative_distribution}
\end{figure}

We find, that in the max-flow model, just as in the electrical
conductance case, scale-free networks exhibit a power-law decay of
the distribution of conductances with the same exponent 
(and thus very high conductance values are
possible), while in Erd\H{o}s-R\'{e}nyi networks, the conductance
decays exponentially (Fig.~\ref{flow_cumulative_distribution}(a)).
In order to better understand this behavior, we plot the
scaled-flow $G_{\rm MF}/k_B$ as a function of the scaled-degree
$x \equiv k_A/k_B$ (Fig.~\ref{flow_cumulative_distribution}(b)). It
can be seen that the transition at $x=1$ is sharp. For all $x < 1$
($k_A < k_B$), $G_{\rm MF}=x$ (or $G_{\rm MF} = k_A$), while
for $x > 1$ ($k_B < k_A$), $G_{\rm MF}=1$ (or $G_{\rm MF} =
k_B$). In other words, the conductance simply equals the minimum
of the degrees of $A$ and $B$. In the symbols of Eq. (\ref{Gh}),
this also implies that $c \rightarrow 1$; i.e. scale-free networks
are optimal for transport in the max-flow sense. The derivation
leading to Eq. (\ref{PhiG}) becomes then exact, so that the
distribution of conductances is given again by $\Phi_{\rm
MF,SF}(G_{\rm MF}) \sim G_{\rm MF}^{-(2\lambda-1)}$.

\begin{figure}
\begin{center}
\epsfig{file=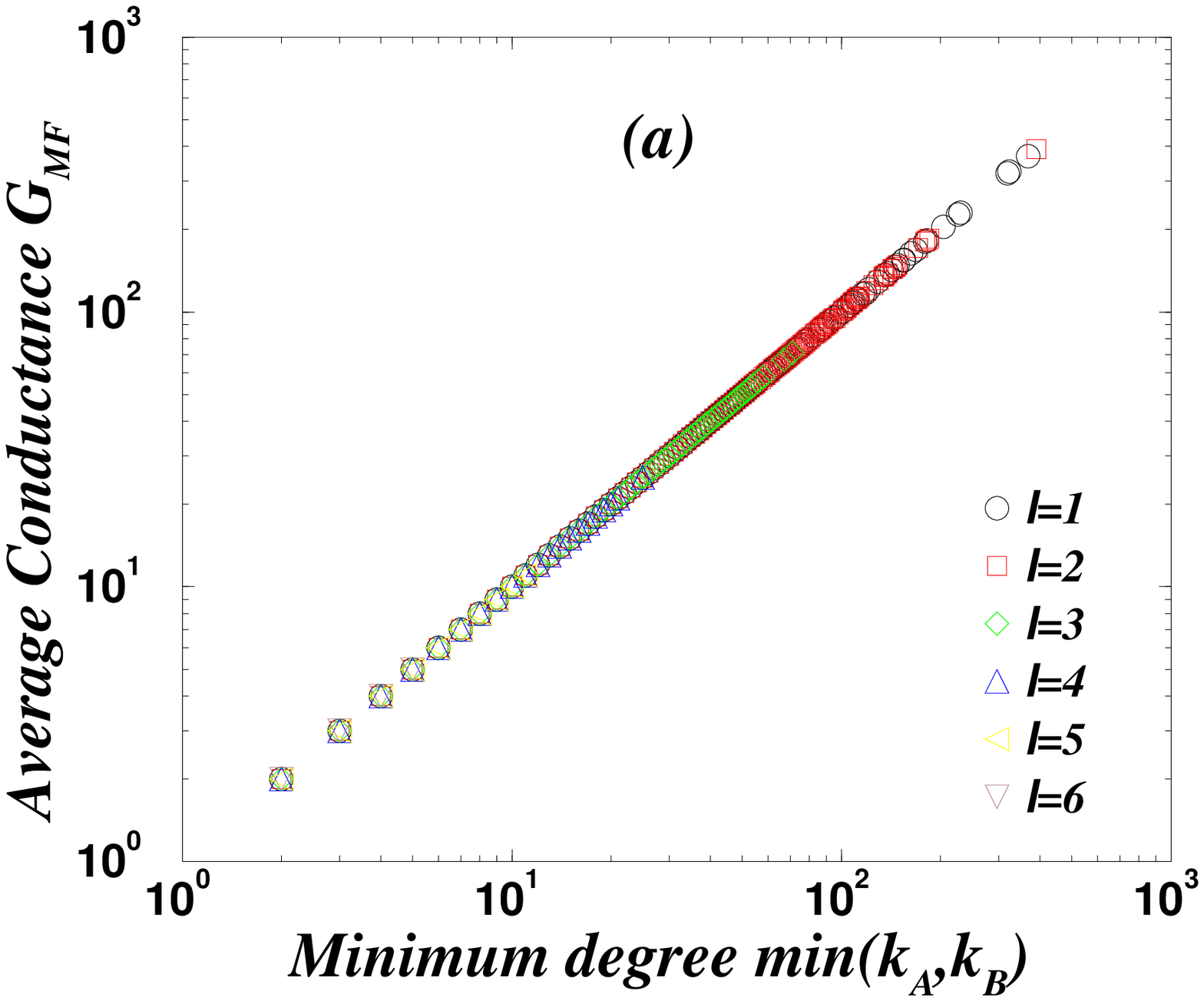,height=6cm,width=6cm}
\epsfig{file=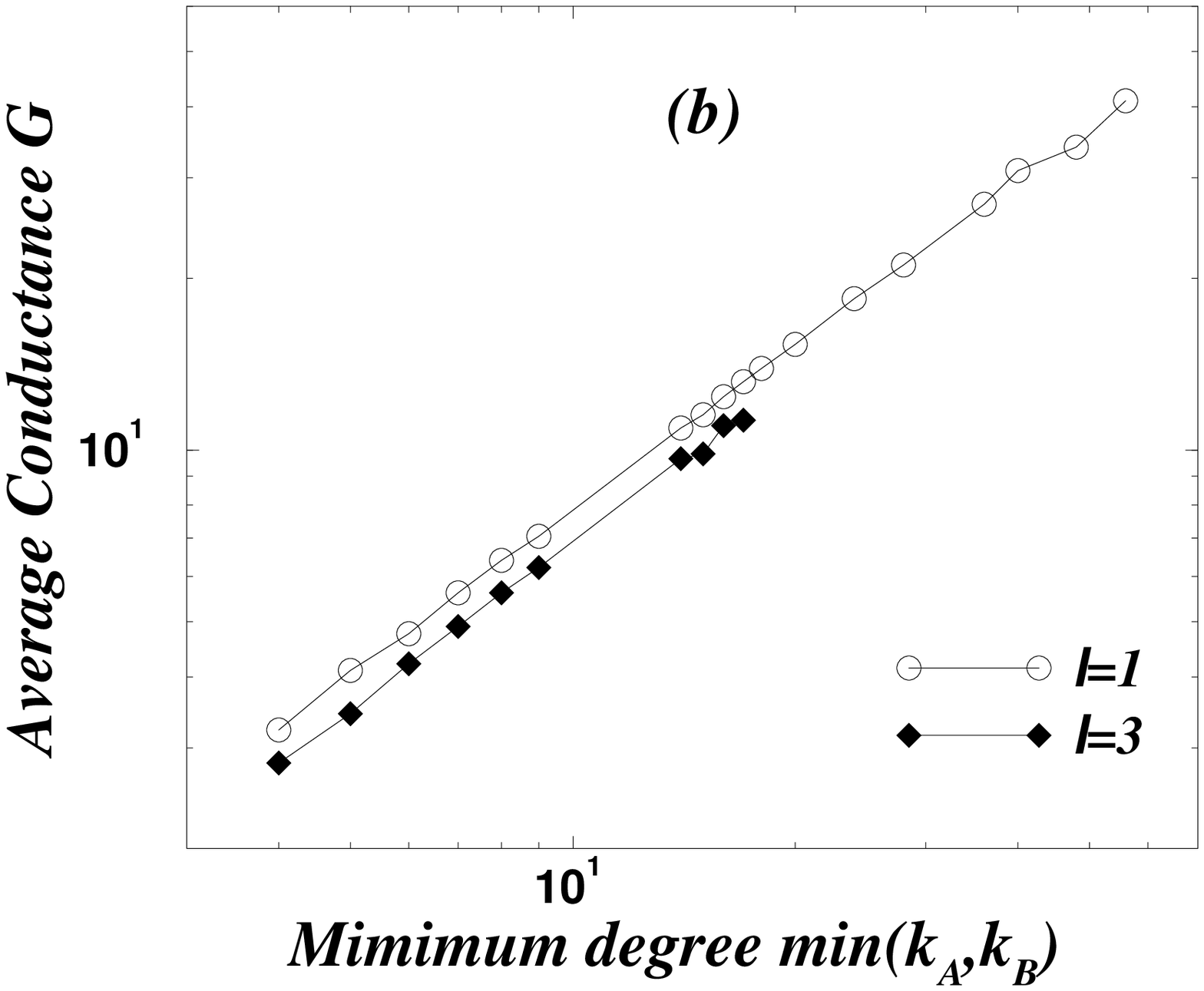,height=6cm,width=6cm}
\end{center}
\caption{(a) Average conductance $\overline{G}_{\rm MF}$ vs.
minimum degree $\min(k_A,k_B)$ of the source and sink $A$ and $B$ for different
values of the shortest distance $\ell_{AB}$. The relation is
independent of $\ell_{AB}$ indicating the independence of
$\overline{G}_{\rm MF}$ on the distance. The network has $N=8000$,
$\lambda=2.5$, $k_{\rm min}=2$. (b) Average conductance $\overline{G}$
vs. minimum degree $\min(k_A,k_B)$ of the source and sink $A$ and $B$ for
different values of distance $\ell_{AB}$. The independence of
$\overline{G}$ with respect to $\ell_{AB}$ breaks down and, as
$\ell_{AB}$ increases, $\overline{G}$ decreases. Once again,
$N=8000$ and $\lambda=2.5$, but the average has been performed for
various $k_B<k_A$ and $k_A=750$.} \label{flow_distance}
\end{figure}

We have so far observed that the max-flow model is quite similar
to electrical conductance, both have similar finite probability 
of finding very high values of conductance. 
Also, the fact that the minimum degree plays a dominant role in the number of
link-independent paths makes the scaling behavior of the
electrical and frictionless problems similar. Only when the
conductances are studied as a function of distance, some
differences between the electrical and frictionless cases begin to
emerge. In Fig,~\ref{flow_distance}(a), we plot the dependence of
the average conductance $\overline{G}_{\rm MF}$ with respect to
the minimum degree $\min(k_A,k_B)$ of the source and sink, for different values of
the shortest distance $\ell_{AB}$ between $A$ and $B$, and find
that $\overline{G}_{\rm MF}$ is independent of $\ell_{AB}$ as the
curves for different $\ell_{AB}$ overlap. This result is a
consequence of the frictionless character of the max-flow problem.
However, when we consider the electrical case, this independence
disappears. This is illustrated in Fig,~\ref{flow_distance}(b),
where $\overline{G}$ is also plotted against the minimum degree
$\min(k_A,k_B)$,
but in this case, curves with different $\ell_{AB}$ no longer
overlap. From the plot we find that $\overline{G}$ decreases as
the distance increases.

\begin{figure}
\begin{center}
\epsfig{file=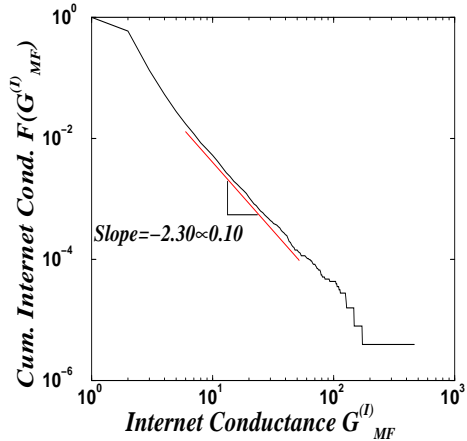,height=6cm,width=6cm}
\end{center}
\caption{Cumulative distribution $F(G^{(I)}_{\rm MF})$ vs $G^{(I)}_{\rm MF}$
  for the Internet. This slope is in good agreement with the scale-free
  structure that has been observed for the Internet (see text).}
\label{flow_internet}
\end{figure}

To test the validity of our model results in real networks, we measured the
conductance $G^{(I)}_{\rm MF}$ on the most up to date map of the
Autonomous Systems (AS) level of the Internet
structure~\cite{Shavitt}. From Fig.~\ref{flow_internet} we find that 
the slope of the plot, which corresponds to $g_G-1$ from Eq.~(\ref{PhiG}), 
is approximately 2.3, 
implying that $\lambda\approx 2.15\pm0.05$, in agreement with 
the value of the degree distribution exponent for
the Internet observed in~\cite{Shavitt}. 
Thus, transport properties can yield information on the topology of
the network.

\section{Multiple Sources}
\label{multiple}

In many cases it is desirable to explore the more general situation,
where the transport takes place between \emph{groups} of nodes,
not necessarily between a single source and a single sink.
This might be a frequent scenario in systems such as the Internet,
where data should be sent between a group of computers to another group,
or in transportation network, where we can consider the quality of transport
between e.g. countries, where the network of direct connections
(flights, roads, etc.) between cities is already established.

The generalization of the above defined models is straightforward.
For the electrical transport case, once we choose our $n$ sources
and $n$ sinks, we simply wire the $n$ sources together to the positive 
potential $V=1$, and the $n$ sinks to $V=0$.
For the max flow case, we connect the $n$ sources to a \emph{super-source}
with infinite capacity links, the same for the sinks. We then 
consider the max flow between the super-source and the super-sink.

For simplicity of notation, in this section we denote the electrical conductance
with $G$ and the max-flow with simply $f$ (instead of $G_{MF}$).

The interesting quantity to consider in this case is the average conductance, or flow,
\emph{per source}, i.e. $\overline{G}/n$ or $\overline{f}/n$ 
(the averaging notation will be omitted when it is clear from the context), 
since this takes into account the obvious
increase in transport due to the multiple sources. Considering the transport
per source gives thus a proper indication of the network utilization.

\begin{figure}
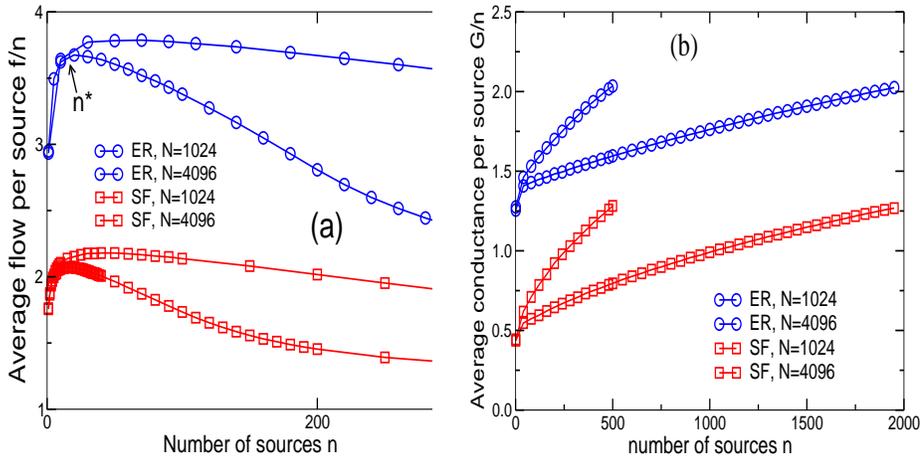

\begin{center}
\epsfig{file=CarmiFig9a.eps,height=6cm,width=6cm}
\epsfig{file=CarmiFig9b.eps,height=6cm,width=6cm}
\end{center}
\caption{(a) The average flow per source, $f/n$ vs. the number of sources $n$,
for ER and SF networks, and two values of network size $N$. For all curves there is an optimal 
$n^*$, for which the flow per source is maximized. Obviously, $n^*$ grows with $N$ 
such that the larger the network is, more users can use it in an optimal way.
Above $n^*$, the flow per source decreases.
(b) The average conductance per source, $G/n$ vs. the number of source $n$
again for ER, SF networks, and different network sizes. Here there is no optimal point--
The more users in the network the more useful it is for transport.
See the text for the qualitative explanation of this behavior.} 
\label{per_source}
\end{figure}

In Fig. \ref{per_source} we analyze the dependence of the transport on $n$.
Consider first the case of max flow, in \ref{per_source}(a). It is seen, that the flow
first \emph{increases} with $n$, and then decreases, for both network models and network sizes.
This observation is consistent with the transport backbone picture described above, 
if we generalize the definition of the transport backbone to include now 
all nodes in the network that are nor sources neither sinks. 
For small $n$ (or more precisely, $n \ll N$), the transport backbone remains as large,
and hence as a good conductor, as it is in the case of a single source ($n=1$).
Therefore, we expect again the flow to depend only on the degrees of the sources and the sinks.
With one source and sink, the flow is confined by the smaller degree of the source and the sink.
With multiple sources and sinks, the flow will be the minimum of \\
$\mbox{[sum of the degrees of the sources, sum of degrees of the sinks]}~$. 
The more sources we have (larger $n$),
the more chances for the sum of the degrees to be larger (since for a sum of many degrees
to be smaller, we need \emph{all} the degrees to be small, which is less probable).
Thus for small $n$, $\overline{f}/n$ increases. An exact theory for this region is developed below.
As $n$ grows larger, not only that the transport backbone becomes smaller, but the number of 
paths that need to go through it grows too. Therefore, many paths that were parallel before
now require the simultaneous usage of the same link, and the backbone is no longer
a perfect conductor. In other words, the \emph{interaction} between the outgoing paths causes
the decrease in the backbone transport capability.

From these two contradicting trends emerges the appearance of an \emph{optimal} $n$, $n^*$.
which is the $n$ for which the flow per source is maximized.
This has an important consequences for networks design, since it tells us
that a network has an optimal ``number of users'' for which the utilization
of the network is maximized. Increasing or decreasing the number of nodes in each group,
will force each node, on average, to use less of the network resources.

In Fig. \ref{per_source}(b), we plot the same for the electrical conductance case.
But, in contrast to the flow case, here the conductance per source only 
\emph{increases} with respect to the number of sources $n$. 
The reason is, that for large $n$, the \emph{number} of parallel paths decrease,
but their \emph{lengths} decrease (since for so many sources and sinks, there is more probability
for a direct or almost direct connection between some source and some sink), and therefore
the conductance of each path significantly increase, such that the total flow increases too.
All this did not affect the flow, since there, as mentioned in 
Section \ref{frictionless}, the total flow do not depend on the distance between the source and 
the sink.

This fact, that electrical conductance improves with more sources, while
flow only degrades, actually points out a fundamental difference between the two types of transport
phenomena. This actually is a consequence of the arguments given in Section \ref{frictionless}
as for the distance dependence, and completes the picture considering the comparison between the models.

For the max-flow model, a closed form formulas for $\overline{f}(n)$ and the
probability distribution of the flow $\Phi_{\rm MF}(f)$ (for a given n) can
be derived analytically in the region where $n \ll N$, where we can assume
there is no interaction at all between the paths of the $n$ pairs, such that
the flow is just the minimum of \\ $\mbox{[sum of the degrees of the source,
sum of the degrees of the sinks]}~$.  A comparison between the theoretical
formula and the simulation results will enable us to directly test this
hypothesis; moreover it will mark the region ($n$ values) where ``no
interaction'' between flow paths exist.

\begin{figure}
\begin{center}
\epsfig{file=CarmiFig10a.eps,height=6cm,width=6cm}
\epsfig{file=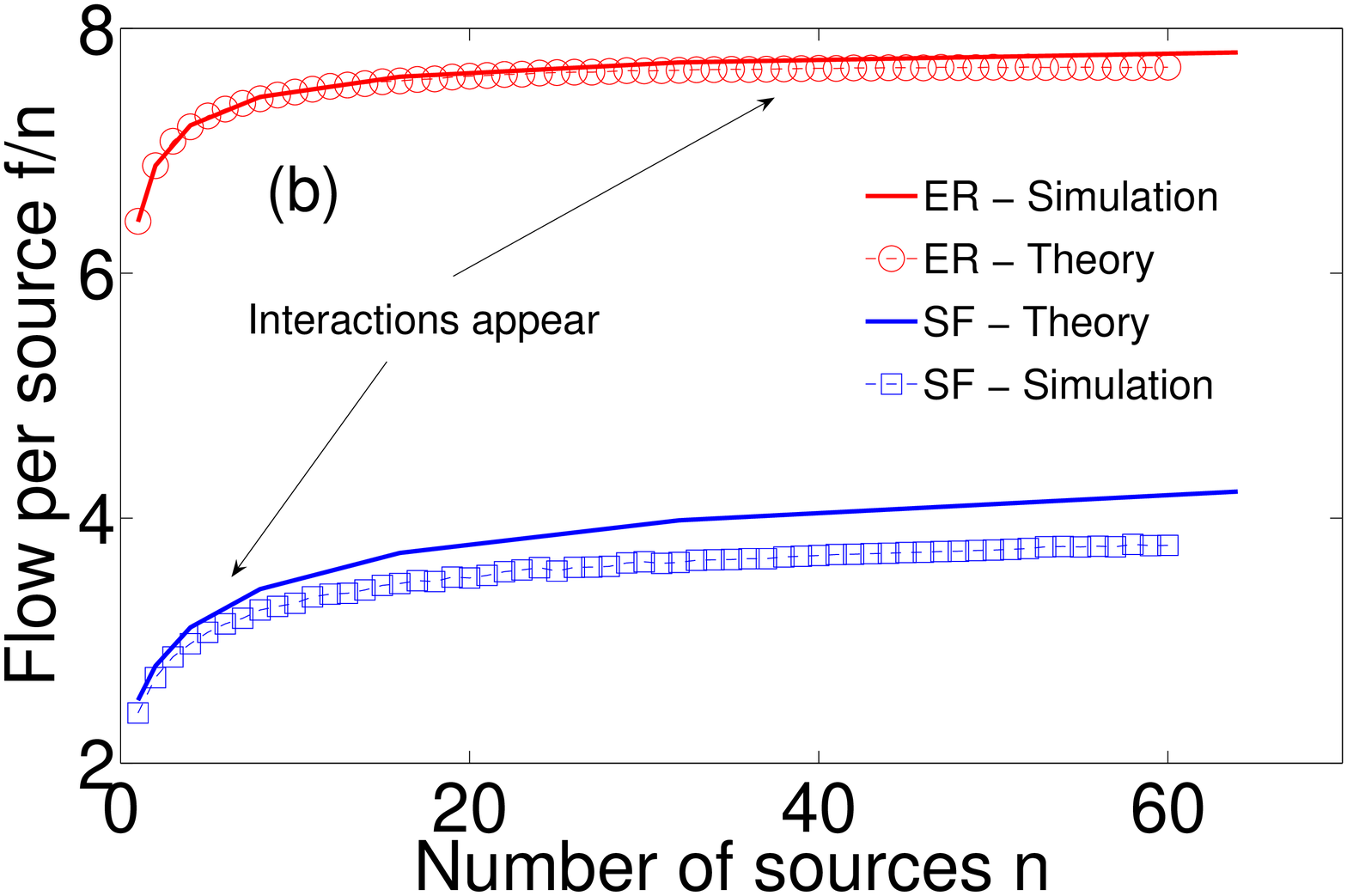,height=8cm,width=8cm}
\end{center}
\caption{(a) For an ER network, with average degree $\overline{k}=4$, $N=8192$, 
we plot the probability distribution of the flows $\Phi_{\rm MF,n}$, for given $n=4,8,16$,
as well as the theoretical prediction from Eq.~(\ref{ER_flow}).
(b) For ER network with $\overline{k}=8$, $N=4096$, and SF network with $\lambda=2.4, k_{\rm min}=2$,
$N=8192$, we plot the average flow per source, $\overline{f}$ vs. $n$,
and compare it with the theoretical prediction under the assumptions of no interactions between 
the parallel paths. The number of sources where this assumption no longer holds is marked.} 
\label{theory}
\end{figure}

Next, we present the analytic derivation for small n. Denote by $k_n$ the sum
of degrees of $n$ nodes.  The total flow between $n$ sources and $n$ sinks is
then given by $f = \min[k_n(\mbox{sources}),k_n(\mbox{sinks})]$.  To
calculate the distribution of flows (from which the average flow is found
easily) one has first to calculate the distribution of the sum of $n$ iidrv's
$k_1$: $P_{k_1}(k) \sim k^{-\lambda}, [k_{\rm min}\le k\le k_{\rm max}]$, for
SF graphs, or $P_{k_1}(k) = \frac{\left(\overline{k}\right)^k
e^{-\overline{k}}}{k!}$ for ER graphs.  For $n \gg 1$, this is easy with the
central limit theorem. However, for very large $n$, the effect of the
interactions will begin to appear. Therefore, one should calculate the
distribution of the sum directly, using convolution.  If we consider the SF
degree distributions to be discrete, The pdf of a sum of degrees of 2 nodes
is then given by -
\begin{equation}
\label{e_sum}
 P_{k_2}(k) = \sum_{j=k_{min}}^{k-k_{min}} P_{k_1}(j) \cdot P_{k_1}(k-j) ~~~;~~~
 2k_{min} \leq k \leq 2k_{max},
\end{equation}
which can be computed exactly in a computer, for a given $N,k_{min},\lambda$. 
The calculation can be easily extended to $n$ equals other pairs of 2.
For ER networks, it is known that the sum of $n$ Poisson variables with parameter $\lambda$
is also a Poisson variable with parameter $n\lambda$.
The probability distribution of the flow itself is readily calculated as the pdf
of the minimum of the $k_n$'s, analogous to Eq.~(\ref{k_min_eq})--
\begin{equation}
\label{e_min} 
\Phi_{\rm MF,n}(f) = 2 \cdot P_{k_n}(f)
\cdot \left[\sum_{j=f}^{\infty} P_{k_n}(j)\right] -
\left[P_{k_n}(f)\right]^2.
\end{equation}
This can also be computed exactly in a computer, together with the average flow 
($\overline{f}(n) = \sum_{f=1}^{\infty} f \cdot \Phi_{\rm MF,n}(f)$).
However, for ER networks the sum can be solved:
\begin{equation}
\label{ER_flow}
\Phi_{\rm MF,n}(f) = 2 \frac{(n \lambda)^{f} e^{-n
\lambda}}{f!} \left( 1 - \frac{\Gamma(f,n\lambda)}{\Gamma(f)} -
\frac{1}{2}\frac{(n \lambda)^{f} e^{-n \lambda}}{f!} \right),
\end{equation}
where $\Gamma(a)$ is the Gamma Function and $\Gamma(a,x)$ is the
Incomplete Gamma Function.
In Fig. \ref{theory}, the ``no-interactions'' theory is compared to the simulation
results, for small $n$. In panel (a), the pdf of the flows is shown for an ER network.
The agreement between the theory and simulation is evident. In panel (b),
we show the average flow per source, $\overline{f}/n$, as computed using Eq.~(\ref{e_min}),
together with the simulation results, for ER and SF networks. The range of applicability
of the no-interactions assumptions is clearly seen.

\begin{figure}
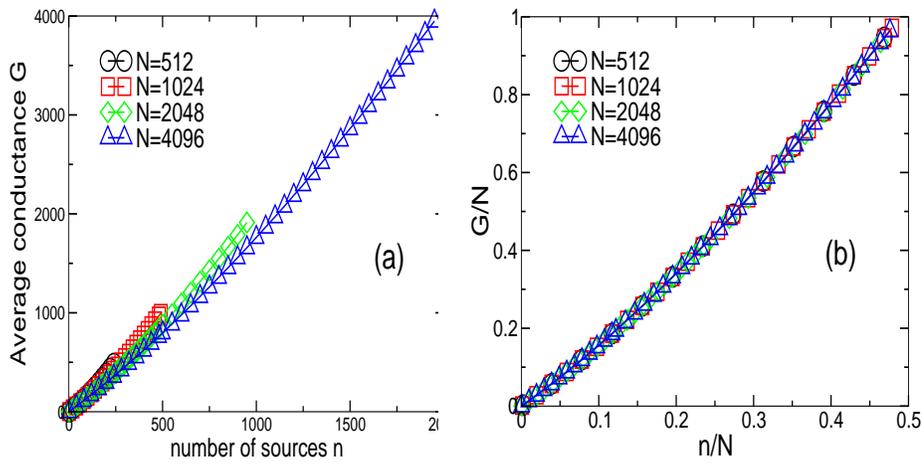

\begin{center}
\epsfig{file=CarmiFig11a.eps,height=6cm,width=6cm}
\epsfig{file=CarmiFig11b.eps,height=6cm,width=6cm}
\end{center}
\caption{(a) The average conductance $\overline{G}$ is plotted vs. the number of sources $n$ for 
ER networks with average degree $\overline{k}$=4, and various values of network sizes $N$. 
(b) The same data, plotted now as $G/N$ with respect to $n/N$.
With this scaling all the curves collapse, implying the scaling law $\overline{G} \sim Ng(n/N)$.} 
\label{scaling}
\end{figure}

Scaling law for the conductance with multiple sources, is seen when properly scaling
the conductance and the number of sources. Plotting $\overline{G}/N$ vs. $n/N$, (Fig. \ref{scaling}),
all curves with different values of $N$ collapse.
Thus we conclude the scaling form:
\begin{equation}
\label{scaling_eq}
\overline{G} \sim N g(n/N),
\end{equation}
where $g(x)$ is a function of a single variable only. The same scaling appears for the flow
as well (data not shown).

\section{Summary}
\label{summary}

In summary, we find that the conductance of scale-free networks is highly
heterogeneous, and depends strongly on the degree of the two nodes $A$ and
$B$. We also find a power-law tail for $\Phi_{SF}(G)$ and relate the tail
exponent $g_G$ to the exponent $\lambda$ of the degree distribution
$P(k)$. This power law behavior makes scale-free networks better for
transport. Our work is consistent with a simple physical picture of how
transport takes place in scale-free and Erd\H{o}s-R\'{e}nyi networks, which
presents the 'transport backbone' as an explanation to the fact that the
transport virtually depends only on the smallest degree of the two
nodes. This scenario appears to be valid also for the frictionless transport
model, as clearly indicated by the similarity in the results. We analyze this
model and compare its properties to the electrical conductance case. We also
compare our model results on a real network of the AS Internet and obtain
good agreement. We then extend the study to transport with multiple sources,
where the transport takes place between two groups of nodes. We find that
this mode arises a fundamental difference between the behavior of the
electrical and frictionless transport models. We also find an optimal number
of sources in which the transport is most efficient, and explain its origin.

Finally, we point out that our study can be further extended.  We could find
a closed analytical formula, Eqs.~(\ref{e_min}),~(\ref{ER_flow}) for the
distribution of flows in ER networks only.  It would be useful to find an
analytical formula also for SF networks, and in addition, to find a
analytical form for the average flow (and conductance), per source.  In other
words, we need to find the scaling function $g(x)$ in Eq. (\ref{scaling_eq}).
Also, it is of interest to investigate the dependence of the optimal number
of sources, $n^*$, in the network size and connectivity. Finally, other,
maybe more realistic, models for transport in a network with multiple sources
should be considered.

\section{Acknowledgments}
We thank the Office of Naval Research, the Israel Science Foundation,
the European NEST project DYSONET, and the Israel Internet Association
for financial support, and
L. Braunstein, R. Cohen, E. Perlsman, G. Paul, S. Sreenivasan,
T. Tanizawa, and S. Buldyrev for discussions.



\newpage

\newpage


\begin{thebibliography}{00}
\bibitem{netwcomment} In principle, a node can have a degree up to
$N-1$, connecting to all other nodes of the network.  The results
presented here correspond to networks with upper cutoff $k_{\rm
max}=k_{\rm min}N^{1/(\lambda-1)}$ imposed.  We also studied networks
for which $k_{\rm max}$ is not imposed, and found no significant
differences in the pdf $\Phi_{\rm SF}(G)$.
\bibitem{dyn-topology} The dynamical properties we study are
related to transport on networks and differ from those which treat the
network topology itself as evolving in time \cite{scale-Barabasi,dyn-network}.
\end{thebibliography}
\end{document}